# A semiconductor exciton memory cell based on a single quantum nanostructure

*Hubert J. Krenner[1$*], Craig E. Pryor[2], Jun He[1] & Pierre M. Petroff[1+]*

[1]*Materials Department, University of California, Santa Barbara, CA 93106, USA*

[2]*Department of Physics and Astronomy, University of Iowa, Iowa City, IA 52242, USA*

[*] krenner@engineering.ucsb.edu, [+] petroff@engineering.ucsb.edu

[$] Corresponding author

Abstract:

We demonstrate storage of excitons in a single nanostructure, a self-assembled Quantum Post. After generation electron and holes forming the exciton are separated by an electric field towards opposite ends of the Quantum Post inhibiting their radiative recombination. After a defined time such spatially indirect excitons are reconverted to optically active direct excitons by switching the electric field. The emitted light of the stored exciton is detected in the limit of a single nanostructure and storage times exceeding 30 msec are demonstrated. We identify a slow tunneling of the electron out of the Quantum Post as the dominant loss mechanism by comparing the field dependent temporal decay of the storage signal to models for this process and radiative losses.

Main text:

The zero-dimensional density of states of self-assembled semiconductor quantum dots (QDs) provides the basis for numerous applications for novel optoelectronic devices such as optical modulators and low-threshold lasers[1] and also allows studying fundamental physical phenomena in these nanostructures. In particular, spins[2,3,4] and excitons[5,6,7,8] have been of interest for implementing quantum computation and communication schemes[9] in the solid state. Indeed, self-assembled QDs are inherently scalable[10,11] and can be addressed, manipulated and read out both electrically and optically[3,4,6,12]. Recently, memory devices based on QDs[3,13,14] have been realized and spin to photon conversion has been demonstrated[3]. However, in these experiments one carrier initially forming the exciton is lost to a reservoir, resulting in the loss of phase information. Here we report the storage of optically or electrically generated excitons in a novel type of nanostructure: a self-assembled quantum post (QP)[15]. Unlike in QD based approaches, in our system both electrons and holes remain localised within the same nanostructure in which they are generated, maintaining the phase between the electron and hole. By applying an electric field along the QP axis electron and hole are spatially separated towards opposite ends of the QP. This results in a large electrostatic dipole moment and a large increase in the radiative lifetime of the exciton. Such stored excitons can be reconverted into photons simply by switching a gate voltage after a defined time. The storage time exceeds 30 ms at 7 K and is only limited by electron tunnelling out of the QP. These timescales are sufficiently long to study the mutual electron-hole spin coherences and reconversion from stationary to flying photonic qBITs using our storage scheme. In addition, the large and tuneable electrostatic dipole moment of the stored exciton gives rise to pronounced change of the dielectric properties and tailored optical non-linearities. This could find direct application in optical phase modulation devices with superior performance.

The novel nanostructure used in these experiments is a self-assembled quantum post[15,16] which is a short quantum wire with a QD at each end and is embedded in a semiconductor matrix. In contrast to well-established strain induced QDs the height of the QP can be controlled with nanometre precision up to 60 nm and the aspect ratio can be engineered. The QPs are embedded in a shallow quantum well preserving confinement in the lateral direction. A detailed description of the QP fabrication and the device can be found in the supporting information. In contrast to quantum wells[17,18] in our system individual excitons are fully localized within a single nanostructure and strongly decoupled from the environment. The controllable height and large aspect ratio allows for charge separation along the QP axis and while keeping the electron and hole localized *within the same nanostructure*. This property is required for the generation and phase preserving storage of indirect excitons and their reconversion into photons which we demonstrate in this paper. Unlike single or coupled QD based approaches, our scheme should allow for full transfer of quantum information from a stationary spin or exciton back into the optical domain of a flying photonic qBIT[19].

We begin with introducing the fundamental optical properties of QPs. In Fig. 1 (a) we show a typical micro-photoluminescence (PL) spectrum recorded from an individual 40 nm high QP under low optical excitation. This QP is not embedded in an electrically active structure and, therefore, no electric field is applied. The spectrum shows three dominant lines corresponding to the QP exciton ground state shell which can be attributed to recombination of the neutral exciton ($X^0 = 1e + 1h$, marked in red) and two charged excitons ($X^+ = 1e + 2h$ and $X^- = 2e + 1h$). The energy splitting between these different charge states is comparable to self-assembled quantum dots[11,16] (e.g. $X^-$ is shifted by 4.25 meV to lower energy compared to $X^0$). We recorded the autocorrelation function $g^{(2)}(\tau)$ of the $X^0$ line under pulsed excitation (inset of Fig. 1 (a)). In the typical periodic pattern the peak at $\tau = 0$ s measures the probability for emission of two photons per excitation cycle. It is clearly suppressed well below the average height of the other peaks marked by the shaded gray bar providing direct evidence that QPs are efficient single photon emitters[7]. In particular, a single quantum emitter is required for efficient conversion between exciton or spin excitations and photons i.e. between stationary and flying qBITs[19].

In a second experiment we studied the influence of a static electric field ($F$) on the QP emission. The underlying tuning mechanism is the Quantum confined Stark effect (QCSE) which has proven to be a particularly powerful tuning mechanism to manipulate, read, or couple qBITs in quantum dots[5,10,11,12]. Fig. 1 (b) shows micro-PL spectra of an individual 40 nm QP encoded in the colour scale as a function of the wavelength and the gate potential ($V_{Gate}$) applied to our device (see supporting information). $F$ is oriented along the QP axis and increases with decreasing $V_{Gate}$ as shown schematically in Fig. 1 (b). For low electric fields ($V_{Gate} \sim 1.4$ V) we observe two dominant spectral lines which originate from recombination of neutral and charged excitons in the QP. As $F$ increases all lines show a weak shift due to the QCSE up to $V_{Gate} = \sim 0.8$ V. At this voltage the PL energy rapidly changes and its intensity quenches. Tunneling of carriers is not expected at such fields ($F \sim 30$ kV/cm), and cannot explain the change in the QCSE. The magnitude of the Stark shift is given by $\Delta E_{Stark} = p \cdot \Delta F$, where $p$ is the static excitonic dipole moment. It is directly related to the mean distance of the electron and hole wavefunctions $s_{e\text{-}h}$ via $p = e \cdot s_{e\text{-}h}$ with $e$ being the elementary charge[10]. For $V_{Gate} > 0.9$ V the electron and hole are strongly bound via their mutual Coulomb interaction and, therefore, $s_{e\text{-}h}$ remains small and almost constant giving rise to a weak shift rate. For $V_{Gate} < 0.9$ V the Stark shift $\Delta E_{Stark}$ becomes the dominant energy contribution and exceeds the binding energy. Therefore, the electron and hole are separated by the electric field and their distance increases, resulting in a larger value of $p$ which is limited by the height of the QP. This increase in $p$ results in a much larger slope of the Stark-shift and a quenching of the PL intensity due to a drastic decrease in oscillator strength. This transition is more abrupt for larger QP height and the slope of the indirect exciton is not resolvable. However for shorter QPs this asymptotic linear shift is observed[16] and an example of a 23 nm high QP is included as supporting Fig. S1. The resolved transition to a linear slope directly reflects the spatial separation of electron and hole. Moreover, the electrostatic dipole moment of $p = e \cdot 20$ nm determined from the slope reflects the QP height. The linear shift reflecting the QP height, the reduction of the oscillator strength and the lateral carrier confinement are due to the unique electronic properties of QPs which in contrast to conventional quantum well or QD structures can be accurately tailored. Furthermore, these two observations provide a clear, characteristic fingerprint for an electric field-driven separation of individual electrons and holes *within* the QP leading to a spatially indirect exciton whose radiative lifetime is substantially and controllably increased.

This unique QCSE enables us to implement a scheme for storing an indirect exciton in a QP as shown schematically in Fig. 2 (a). Our method uses a sequence of three electrical pulses for gating: the sample, the laser to write excitons, and the detector to read the stored signal. For each step the spatial distribution of the electron and hole and the band structure are depicted schematically. The storage sequence starts with a reset step under a large reverse bias ($V_{reset} \sim$ -5 V) under which carriers can tunnel out of the QP. In the next step, the sample bias is raised to $V_{store}$, a level where the exciton is in a spatially indirect long-lived state while carrier tunnelling out of the QP is suppressed. At the start of this storage interval, the laser is turned on for 500 ns to write photogenerated electrons and holes into the QPs. During storage both carriers are localized *in the same* nanostructure but remain spatially separated due to the applied electric field. When the sample bias is brought back close to flat-band ($V_{read}$) for a time $t_{read} = 1$ μs, the stored excitons become spatially direct. Under these conditions the electron and hole wavefunctions overlap and radiative recombination can occur. The emitted photons are then detected during this read step. To reduce the dark count level the detector is read out only while the read voltage is applied. Thus, we define the storage time ($t_{store}$) as the difference between the time when the "write" laser is turned off and the time a "read" voltage is applied.

To demonstrate this scheme we used a sample with a low surface density of QPs ($\sim$1000/μm$^2$). The store and read voltages were set to $V_{store} = +0.5$ V and $V_{read} = +1.5$ V, respectively. Typical storage spectra as a function of wavelength are shown in Fig. 2 (b) for $t_{store} = 20$ μs and 220 μs. We find a pronounced peak centred at $\lambda = 987$ nm with a width of 34 nm consistent with the QP ground state PL for a QP ensemble[15]. The intensity of this storage peak does not decrease when the storage time is increased by more than an order of magnitude demonstrating loss-free storage of excitons. Higher excitation densities are required for storage compared to PL experiments (e.g. shown in Fig. 1) since at these voltages the carrier capture is significantly quenched. The levels used were chosen a factor $\sim$ 10 higher than the excitation density at which the onset of the storage signal is observed. We want to note that because of the non-resonant nature of the excitation, we cannot exclude that charged excitons are contributing to the storage signal. Since the excitation densities are not significantly increased beyond the observed onset multi-excitons (bi- or triexcitons) should still be negligible.

Two control experiments were performed to demonstrate that we are indeed storing an indirect exciton. In the first one, the read voltage is set to a level at which the exciton ground state remains indirect. For this condition, we do not expect to detect a stored signal since radiative recombination strongly suppressed for indirect excitons. To confirm that the read signal originates from excitons photogenerated by the write laser and not from carriers electrically injected from the doped layers, a second control experiment was done in which the laser remains off. No storage signal should be detected even though the read voltage is applied since no excitons were written. The results of these two control experiments are shown as blue (no read) and green (no write) symbols for $t_{store}$ = 20 μs in Fig. 2 (b). Clearly, no signal is detected in both cases demonstrating that we store excitons in the QPs, which are later read-out simply by changing the gate potential on our device.

In another experiment we extended our storage scheme down to the limit of a single QP. We recorded storage signals of $1.1 \pm 0.1 \cdot 10^{-4}$ and $4.8 \pm 0.5 \cdot 10^{-4}$ counts/cycle, after storage times of 20 μs and 30 μs, respectively, from two individual QPs (spectra shown in supporting Fig. S2). Since only a single QP is addressed a reduced excitation power densities ($P_{exc}$ = 100 W/cm$^{-2}$) is used. To avoid monochromator losses the emission of the QPs was detected after spectral filtering using a 10nm bandpass filter centered at the single QP main emission line as shown in supporting Fig. S2. In the control experiments without the write or read step we measured a background of $0.7 \pm 0.1 \cdot 10^{-4}$ and $2.0 \pm 0.5 \cdot 10^{-4}$ counts/cycle for the two QPs. From these values we obtain net storage signals of $0.4 \pm 0.2 \cdot 10^{-4}$ and $2.8 \pm 1 \cdot 10^{-4}$ counts/cycle. These values agree well with the expected, instrumentation limited signal strength from a single QP of $\sim 10^{-4}$ counts/cycle. The higher count rate for 30 μs storage time compared to 20 μs is associated to differences in the alignment which are most pronounced at these low count rates. As shown in Fig. 2 losses do not occur for times exceeding 200 μs and, therefore, can be excluded. The detected contrast between the full storage scheme and the control experiments clearly demonstrates that even the weak storage signal of a single QP can be detected even without using microcavity structures.

In Fig. 3 (a) we plot the integrated storage signal as a function of $t_{storage}$ for three different values of $V_{storage}$. For $V_{storage} = +0.1$ V ($F \sim 50$ kV/cm) we observe a slow exponential decay of the storage signal with a time constant of 30±7 ms. As $V_{storage}$ is lowered and the electric field is increasing we observe faster decays with 7±1 ms and 0.7±0.2 ms for 0.0V ($F \sim 54$ kV/cm) and -0.1V ($F \sim 58$ kV/cm), respectively. For the radiative lifetime we expect the opposite voltage dependence since the increasing separation of electron and hole reduces their wavefunction overlap. This is confirmed by calculations of the radiative lifetime of spatially indirect excitons as a function of $F$ for a realistic QP morphology using a strain-dependent 8-band $k \cdot p$ model[15,16,20]. They are shown by the black line in Fig. 3 (b) and for a larger range of $F$ in the supporting Fig. S3. However, the observed lifetime shortening is well explained by a reduction of the tunnelling time between the confined QP states and the GaAs matrix schematically shown in Fig. 3 (a). In a simple model the tunnelling time ($t_{tunnelling}$) can be described by a 1-dimensional *WKB* approximation (Fowler-Nordheim tunnelling) given by

$$1 \big/ t_{tunnelling} = \frac{\hbar \pi}{2 m_e^* L^2} \cdot \exp\left[\frac{-4}{3 h e F} \sqrt{2 m_e^* E_{ion}^3}\right] \ (1).$$

Here, $m_e^*$, $L$ and $E_{ion}$ are the electron effective mass, the length of the confinement potential and the ionization energy, respectively. We assume that electron tunnelling is faster than that of holes due to their smaller effective mass[21] and compare in Fig. 3 (b) $t_{tunnelling}$ (red line) with the experimentally measured decay times. The experimental data is well reproduced by this model using $m_e^* = 0.062 \, m_0$ and $E_{ion} = 170$ meV which are in good agreement with our calculations and PL data[15,16]. The radiative contribution shows the opposite behaviour as the tunnelling process and decreases with increasing field due to a reduction in the electron-hole overlap. At $\sim 50$ kV/cm the timescales for tunnelling and radiative losses are comparable giving rise to the maximum storage times of 50 ms. Since our experimental data is reproduced by the computed behaviour for tunnel escape of electrons, we conclude that this process is the dominant loss mechanism in this field range. Clearly, the storage time could be further increased to seconds, limited only by radiative decay (Figure S3) by suppressing tunnel escape with a higher bandgap material like AlGaAs for capping the QPs.

In a final experiment we investigated the temporal statistics of the storage signal. By moving our detection time window relative to the sample and laser sequence we found that the storage signal is only detected during the read voltage pulse as shown in supporting Fig. S4. To improve our temporal resolution we performed a time-correlated single photon counting experiment of the storage signal. In Fig. 4 the logarithmic number of detection events is plotted as a function of the delay $\Delta t$ between the time the read voltage is turned on and the time the reconverted photon is detected. For the storage signal we find a narrow peak with a width of 15 ns and rise and fall times of $t_{rise} = 3.5$ ns and $t_{fall} = 5.0$ ns, respectively. No more photons are detected afterwards over the duration ($t_{read} = 1$ μs) of the read voltage pulse as shown in the inset. We want to note that this fast read-out excludes carrier injection from the doped reservoirs which are separated from the QPs by 110 nm intrinsic GaAs barriers, confirming storage of electrons and holes within the QPs. Moreover, this fast read-out is not possible with conventional QD based schemes and in our device is only limited entirely by the *RC* time constant of the device itself. It can be optimized for efficient, fast exciton-photon conversion for quantum information implementations or triggered single photon generation.

Our device and storage scheme provide exactly the desired behaviour for applications in coherent and optoelectronic devices since it can be readily extended to purely electrical pumping. Here, carriers are injected electrically from the doped reservoirs under a large forward bias. After fast electrical switching to the storage voltage, radiative recombination is suppressed and the remaining excitons are stored. We find efficient storage for electrical injection and a comparison of two spectra taken under electrical and optical pumping is shown in the supporting Fig. S5. This injection scheme could be crucial for a deterministic, electrically pumped single photon source or refreshing of optical phase modulation devices.

In summary, we have demonstrated a novel scheme to store a single exciton in a single QP, without the dephasing transfer of one carrier into a reservoir, as required in a QD based method. Long storage times of more than 30 ms are found which are not limited by radiative recombination and which are not accessible for pairs of QDs since tunnel coupling is required for coherent spin-preserving inter-dot charge transfer[10,11,14]. By modelling potential physical mechanisms as a function of the applied electric field we identified a slow tunnelling process of electrons out of the QPs which dominates the loss of the stored excitons. Furthermore, the observed timescales exceed measured relaxation and decoherence times of spins in QDs[3,4,22]. Since both electron and hole are kept in the memory their *mutual* spin coherence could be probed directly and without the limitation of a short radiative lifetime. For further detailed studies on individual QPs our device can be combined with optical cavities to increase light extraction [23,24] and significantly improve the signal/noise ratio in these delicate experiments. Since our scheme preserves phase information it could provide a direct route to convert stationary (spin or exciton) into flying (photon) qBITs[19]. This would open a whole field of experiments and novel concepts for implementing optical and electrical quantum computation schemes. To ensure this a resonant[3] pumping scheme of the optically active direct exciton preferentially using $\pi$-pulses[5] is required to prevent charged and multi-excitonic contributions. This could be achieved by reducing the device capacitance by shrinking its dimensions to allow for faster switching. With such a device the electric field can be switched to storage conditions after the exciton is generated using a resonant optical pulse before radiative decay can take place. The electrostatic dipole moments of the optically or electrically generated and stored indirect excitons exceed those of single or coupled QDs by several orders of magnitude and are well suited for applications in electro-optical devices with superior performance[25].

This work was supported by NSF via Nanoscale Interdisciplinary Research Team grant CCF-0507295 and NSEC-Harvard. H.J.K. acknowledges support by the Alexander-von-Humboldt-Foundation. We thank M. T. Rakher for help with the autocorrelation measurements.

Supporting information available: Description of sample growth, device design and fabrication. Six supporting figures

Figure 1 – Characterization of individual QPs (a) Micro-PL spectrum of a single QP with an autocorrelation histogram (inset) for the neutral exciton ($X^0$, marked in red). $T$ =7 K and $P_0$ = 1 W/cm$^{-2}$. As expected for a single photon source the peak at t = 0 ns is clearly reduced compared the average peak height (shaded bar) indicative for a single quantum emitter. (b) Transition from direct to indirect excitons observed in the photoluminescence spectrum of a single QP under applied bias voltage at $T$ =7 K and $P_0$ = 50 W/cm$^{-2}$. For decreasing gate potential the electron-holes pairs get spatially separated by the increase of the applied electric field as shown schematically. At $V_{Gate}$ ~ 0.6 V the carrier overlap vanishes and the emission quenches which is accompanied by a change of the shift rate due to the QCSE. The small inserts correspond to a cross section TEM of a QP with the electron and hole wave functions schematically added to it.

Figure 2 – Storage of indirect excitons in QPs. (a) Schematic of the storage scheme. After a reset step during which the QP is emptied the sample voltage (black) is switched to the indirect exciton regime. In the beginning of the storage interval electrons and holes are written by a laser (red) into the QP and after $t_{storage}$ read via direct excitons. The detection (blue) is activated only during the time the read step. The bandstructure and spatial carrier distribution during each step are shown schematically. For control experiments the write and read pulses are turned off individually as shown by the dashed lines. (b) Storage spectra taken from a QP ensemble after 20 ms (full squares) and 200 ms (line) storage times ($P_{exc}$ = 20 kW/cm$^{-2}$, $T$ = 7 K). Without the write and read pulses (open symbols) no signal is detected proving the storage principle.

Figure 3 – Temporal decay of the storage signal and identification of loss mechanism. (a) For increasing reverse bias (increasing electric field) the storage signal decays over time due to tunnel escape of electrons from the QPs as shown schematically. (b), Comparison between experimentally observed decay times (symbols) and calculated tunnelling time (red line) using a WKB approximation and radiative life time (black line) as a function of F demonstrates that tunnelling is the dominant loss mechanism.

Figure 4 – Fast exciton-photon reconversion. In the time-resolved storage signal photons are detected in a short window after the read voltage is applied (see inset). The stored excitons are efficiently reconverted into photons within 25 ns after reset much shorter than the 1ms duration of voltage pulse. The fast rise time of 3.5 ns is limited by the device impedance and demonstrates direct reconversion from excitons into photons and excludes injection of one carrier species from the doped contact layers.

(1)    Bhattacharya, P.; Ghosh, S; Stiff-Roberts, A. D. *Annu. Rev. Mater. Res.* **2004,** *34,* 1.

(2)    Hanson, R.; Kouwenhoven; L. P., Petta, J. R.; Tarucha, S.; Vandersypen, L. M. K. *Rev. Mod. Phys.* **2007**, *79,* 1217.

(3)    Kroutvar, M.; Ducommun, Y.; Heiss, D.; Schuh, D.; Bichler, M.; Abstreiter, G.; Finley, J. J. *Nature* **2004**, 432, 81-84.

(4)    Greilich, A.; Yakovlev, D. R.; Shabaev, A.; Efros, Al. L.; Yugova, I. A.; Oulton, R.; Stavarache, V.; Reuter, D.; Wieck, A.; Bayer, M. Science **2006**, *313,* 341-345.

(5)    Zrenner A.; Beham, E.; Stufler, S; Findeis, F.; Bichler, M.; Abstreiter, G. *Nature* **2002**, 418, 612–614.

(6)    Xu, X.; Sun, B.; Berman, P. R.; Steel, D. G.; Bracker, A. S.; Gammon, D.; Sham, L. J. *Science* **2007**, *317,* 929-932.

(7)    Michler P.; Kiraz, A.; Becher, C.; Schoenfeld, W. V.; Petroff, P. M.; Zhang, L.; Hu, E.; Imamoglu, A. *Science* **2000**, *290,* 2282-2285.

(8)    Akopian, N.; Poem, E.; Berlatzky, J. A.; Avron, J. ; Gershoni, D. ; Gerardot, B. D. ; Petroff, P. M. *Phys. Rev. Lett.* **2006**, *96,* 130501.

(9)    Cerletti, V.; Coish, W. A.; Gywat, O.; Loss, D. *Nanotechnology* **2005**, *16,* R27-R49.

(10)  Krenner, H. J.; Sabathil, M.; Clark, E. C.; Kress, A.; Schuh, D.; Bichler, M.; Abstreiter, G.; Finley, J. J. *Phys. Rev. Lett.* **2005**, *94*, 57402.

(11)  Stinaff, E. A.; Scheibner, M.; Bracker, A. S.; Ponomarev, I. V.; Korenev, V. L.; Ware, M. E.; Doty, M. F.; Reinecke, T. L.; Gammon D. *Science* **2006**, *311*, 636-639.

(12)  Stufler, S.; Ester, P.; Zrenner, A., Bichler, M. *Phys. Rev. Lett.* **2006**, *96*, 037402.

(13)  Finley, J. J.; Skalitz, M.; Arzberger, M.; Zrenner, A.; Böhm, G.; Abstreiter, G. *Appl. Phys. Lett.* **1998**, 73, 2618-2620.

(14)  Lundstrom, T.; Schoenfeld, W.; Lee, H.; Petroff, P. M. *Science* **1999**, *286*, 2312-2314.

(15)  He, J.; Krenner, H. J.; Pryor, C.; Zhang, J. P.; Wu, Y.; Allen, D. G.; Morris, C. M.; Sherwin, M. S.; Petroff, P. M. *Nano Lett.* **2007**, *7*, 802-806.

(16)  Krenner, H. J.; Pryor, C.; He, J; Zhang, J. P.; Wu, Y.; Morris, C. M.; Sherwin, M. S.; Petroff, P. M. *Physica E*, **2008**, *40*, 1785-1789.

(17)  Polland, H.-J.; Schultheis, L.; Kuhl, J.; Göbel, E. O.; Tu, C. W. *Phys. Rev. Lett.* **1985,** *55*, 2610.

(18)  Golub, J. E.; Kash, K.; Harbison, J. P.; Florez, L. T. *Phys. Rev. B* **1990,** *41*, 8564 - 8567.

(19)  Cirac, J. I.; Zoller P.; Kimble H. J.; Mabuchi, H. *Phys. Rev. Lett.* **1997**, *78*, 3221-3224.

(20)  Pryor, C. *Phys. Rev. B* **1998**, *57*, 7190-7195.

(21)  Fry, P. W.; Finley, J. J.; Wilson, L. R.; Lemaitre, A.; Mowbray, D. J.; Skolnick, M. S.; Hopkinson, M.; Hill, G.; Clark, J. C. *Appl. Phys. Lett.* **2000**, *77*, 4344-4346.

(22)  Gerardot, B. D.; Brunner, D.; Dalgarno, P. A.; Öhberg, P.; Seidl, S.; Kroner, M.; Karrai, K.; Stoltz, N. G.; Petroff, P. M.; Warburton R. J. *Nature* **2008,** *451*, 441.

(23)  Badolato  A.;  Hennessy,  K.;  Atatüre,  M.;  Dreiser,  J.;  Hu,  E.;  Petroff,  P. M.;  Imamoglu,  A. *Science* **2005**, *308*, 1158-1116.

(24)  Stoltz, N. G.; Rakher, M. T.; Strauf, S.; Badolato, A.; Loftgreen, D. D.; Petroff, P. M.; Coldren, L. A.; Bouwmeester, D. *Appl. Phys. Lett.* **2005**, *87*, 031105.

(25)  Moreau, G.; Martinez, A.; Cong, D. Y.; Merghem, K.; Miard, A.; Lemaître, A.; Voisin, P.; Ramdane P.; Krestnikov, I.; Kovsh A. R.; Fischer, M.; Koeth J. *Appl Phys. Lett.* **2007,** *91*, 091118.

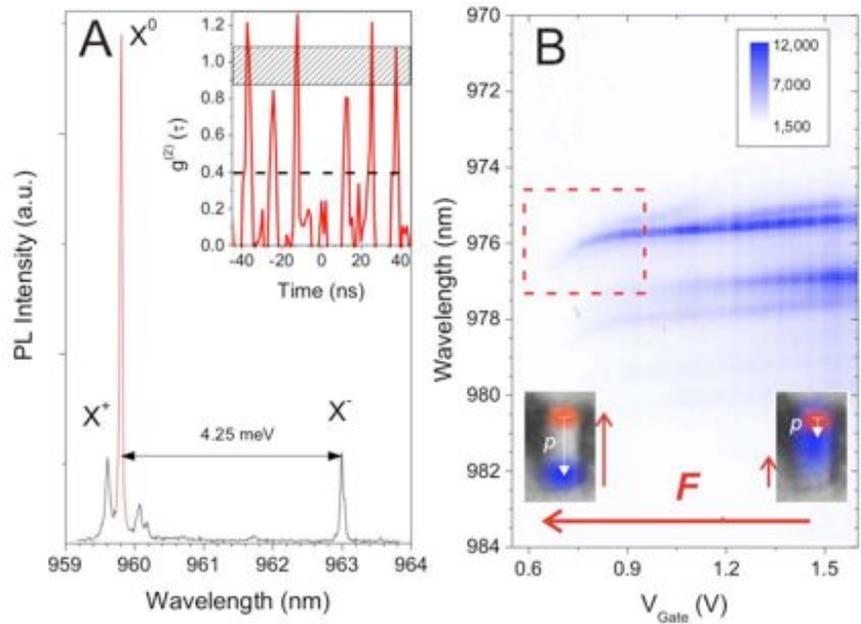

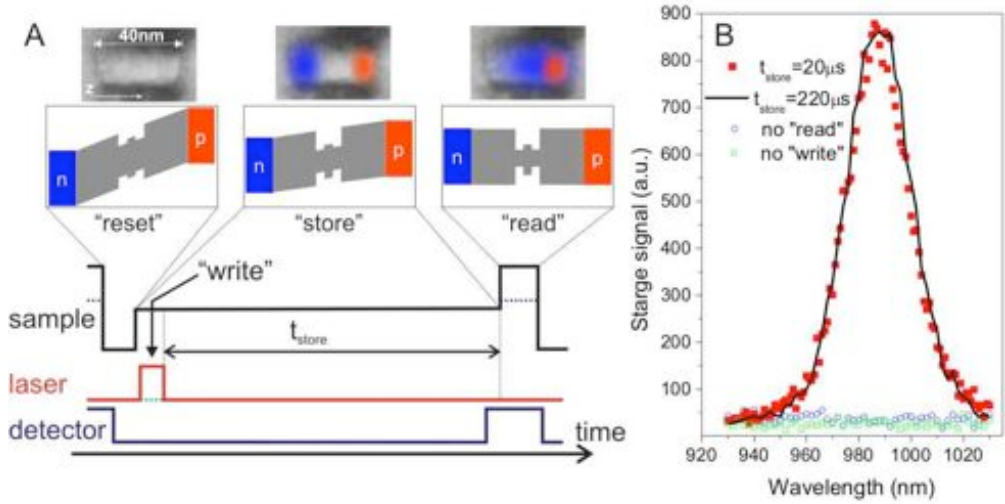

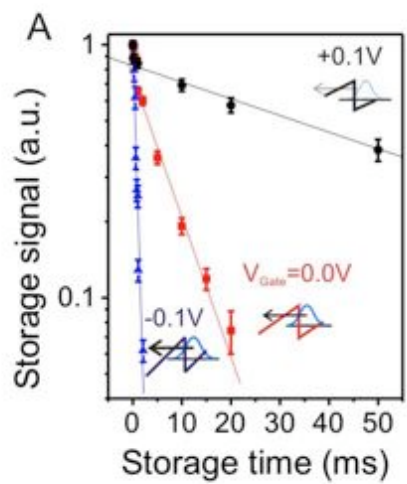

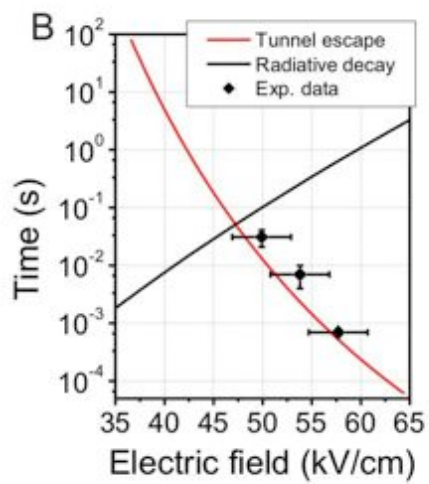

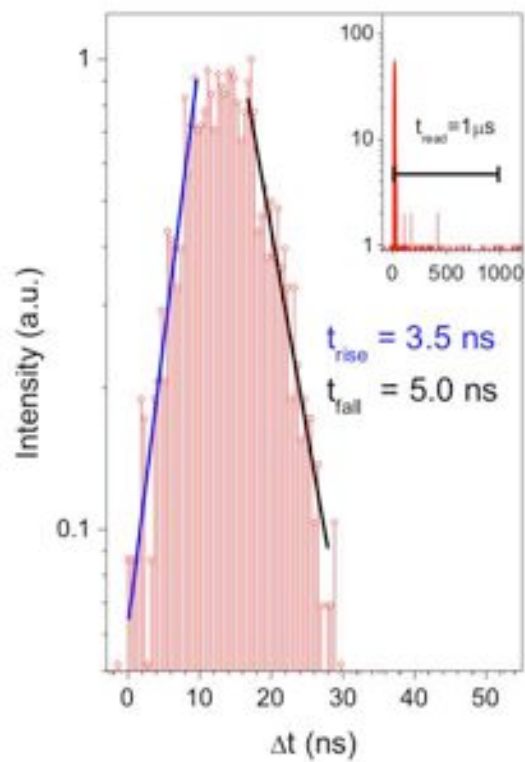

Supporting Information

Sample Growth and Fabrication:

The semiconductor material was grown by solid-source molecular beam epitaxy (MBE) on semi-insulating (001) - GaAs substrates. The epitaxial sequence was as follows: A single layer of 40 nm high self-assembled InGaAs QPs was embedded in the centre of an *n-i-p* junction. The *n*-contact is realized by silicon doped (100 nm $n^+$ = 5·$10^{18}$ cm$^{-3}$, 100 nm $n$ = 1·$10^{18}$ cm$^{-3}$) GaAs which is overgrown by 110 nm of nominally undoped material. For QP growth the substrate temperature is lowered to 530 °C and the rotation of the sample is stopped to realize a surface density gradient across the wafer. The 40 nm high QPs are formed by 16 repetitions of a growth cycle consisting of depositions of 1 monolayer of InAs and 7 monolayers of GaAs each followed by a 60 s growth interruption. The rebuilding and formation of the QPs during the growth interruptions was monitored and confirmed by characteristic changes in the reflection high energy electron diffraction (RHEED) pattern[15]. The QPs were capped by 110 nm of *i*-GaAs to realize a symmetric structure. The sample is finalized by a Beryllium doped *p*-contact (100 nm $p$ = 1·$10^{18}$ cm$^{-3}$, 100 nm $p^+$ = 5·$10^{18}$ cm$^{-3}$). This material was processed into photodiode arrays with a common annealed (GeAuNiAu) *n*-contact on the substrate side and *p*-side (TiPtAu) top contacts on 400 µm square mesas. For experiments on single QPs opaque metal shadowmask patterned with sub-micron apertures were used for optical isolation. The fabricated devices showed rectifying current-voltage characteristics with breakdown voltages at ~ 9 V in reverse and current onset at < 1.5 V in forward direction. This value is in good agreement with the calculated built-in voltage of 1.4 V which we used as the origin of the electric field scale. The QP ensemble was characterized using voltage-dependent micro-PL to determine the bias levels for storage experiments (supporting Fig. S6). For autocorrelation measurements we used a control sample without doped layers.

Experimental Techniques:

For optical measurements, the samples were mounted in a liquid He flow cryostat with a base temperature of $T$ = 7 K. Carriers were photogenerated by continuous wave or pulsed Titanium-Sapphire lasers ($\lambda$ = 850 nm) or a pulsed diode laser ($\lambda$ = 835 nm) which could be electrically gated for storage experiments. The excitation light was focussed to a ~ 1 µm spot using a high numerical aperture objective. Emission from the sample was collected by the same objective, dispersed by a 0.19 m grating monochromator and detected with a spectral resolution <0.5 meV by a liquid $N_2$ cooled charge coupled device (CCD) detector or an avalanche photodiode (APD). Electrical pulses were provided by a computer card and a variable delay generator with rise times > 5 ns for sample gating. A time to amplitude converter card (TAC) was used to measure the auto-correlation function of the exciton emission and the time-dependence of the storage signal. For storage experiments on single QPs the signal was sent directly to the APD after spectral filtering by narrow-band interference filters in order to avoid losses in the monochromator.

Supporting Figures:

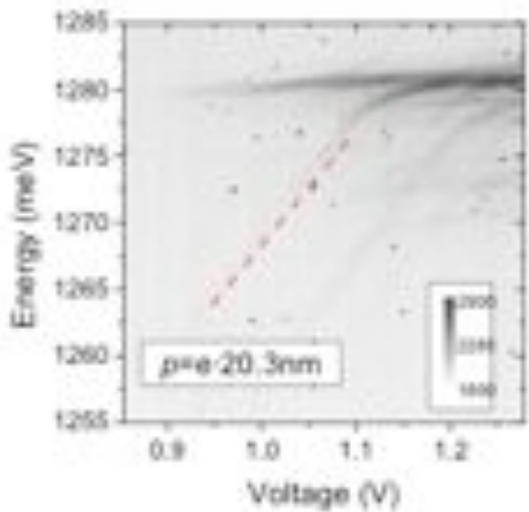

**Figure S1** – PL spectra of a single 23 nm high QPs as a function of the applied gate voltage ($V_G$). In the linear Stark-shift regime from electrostatic dipole moment an electron-hole separation of 20 nm can be determined in excellent agreement with the QP height.

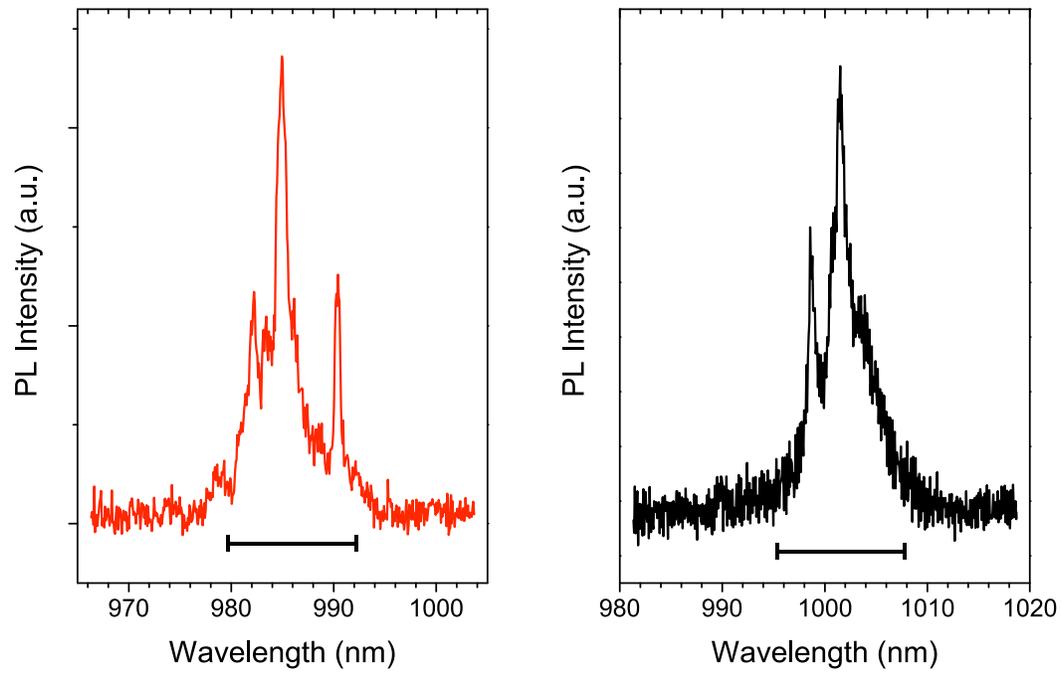

**Figure S2** – PL spectra of two single QPs used for storage experiments recorded at $V_{Gate}$ = 1.5 V. The horizontal line indicates the bandpass of the filter used in for the storage experiments.

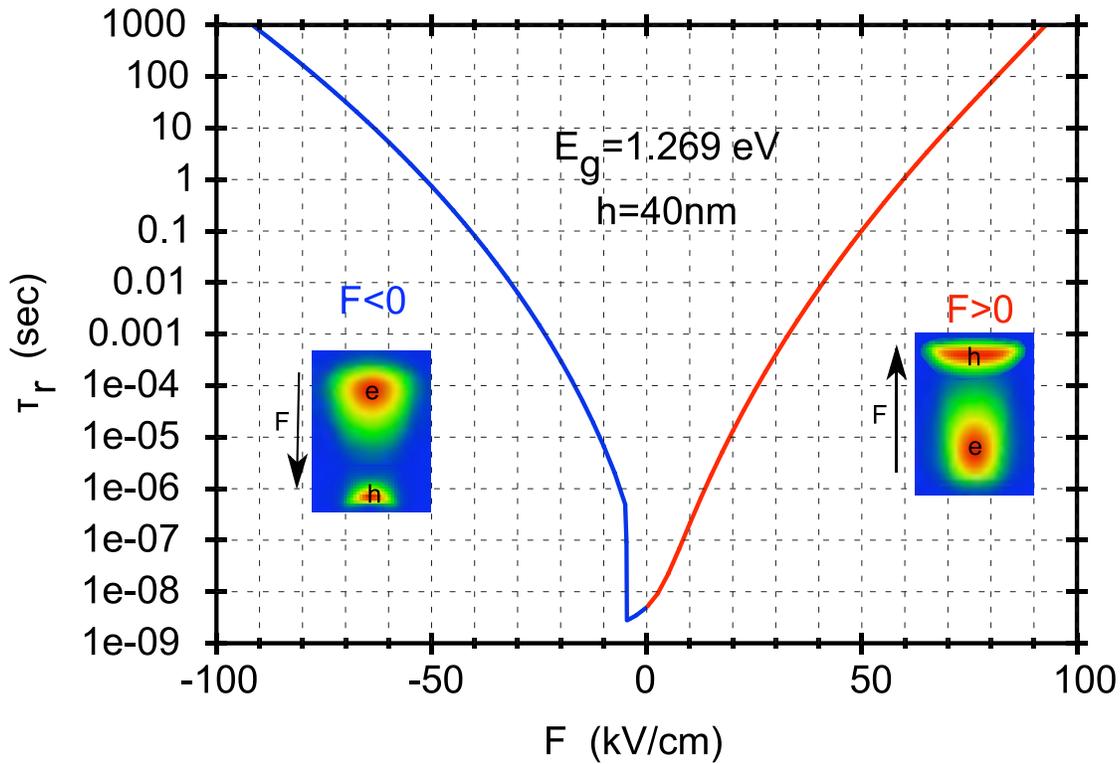

**Figure S3** – Calculated radiative lifetime as a function of the applied axial electric field (*F*). For increasing F the radiative lifetime increases by more than 9 orders of magnitude from ~ 2 ns to several seconds. This behavior is qualitatively the same for both field orientations.

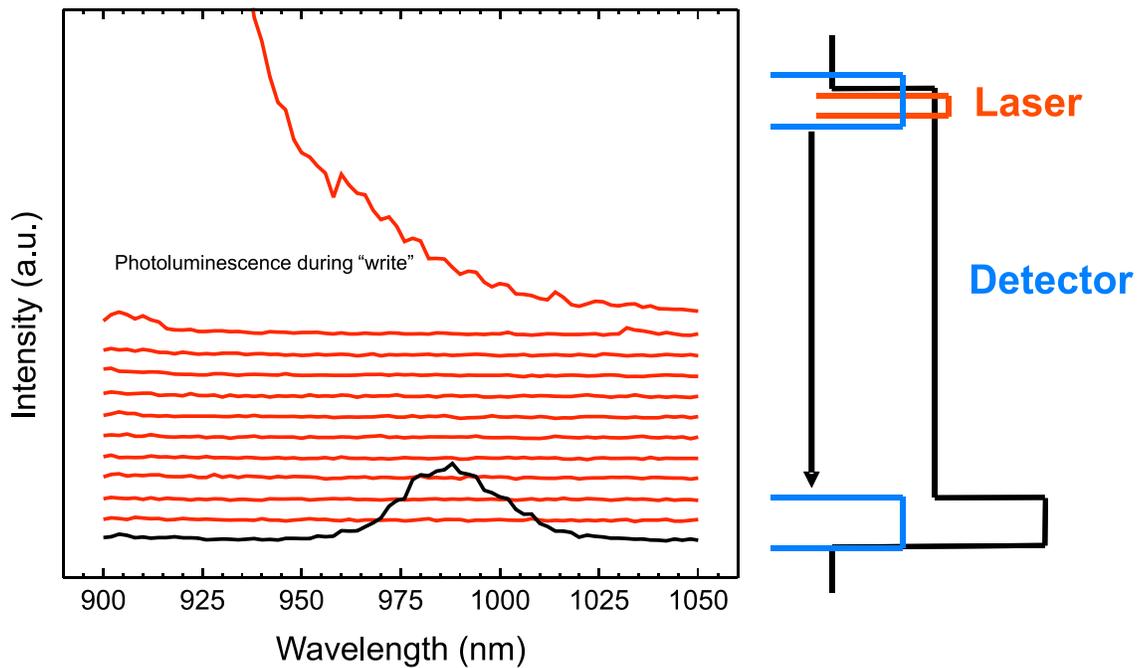

**Figure S4** – Spectra (left panel) recorded with 1 µs detection windows moved in 2 µs steps across the storage cycle as shown schematically in blue on the right panel. A bright PL signal is recorded during the time the "write" laser is on. No signal is detected during the storage time until the "read" voltage is applied on the device (black line).

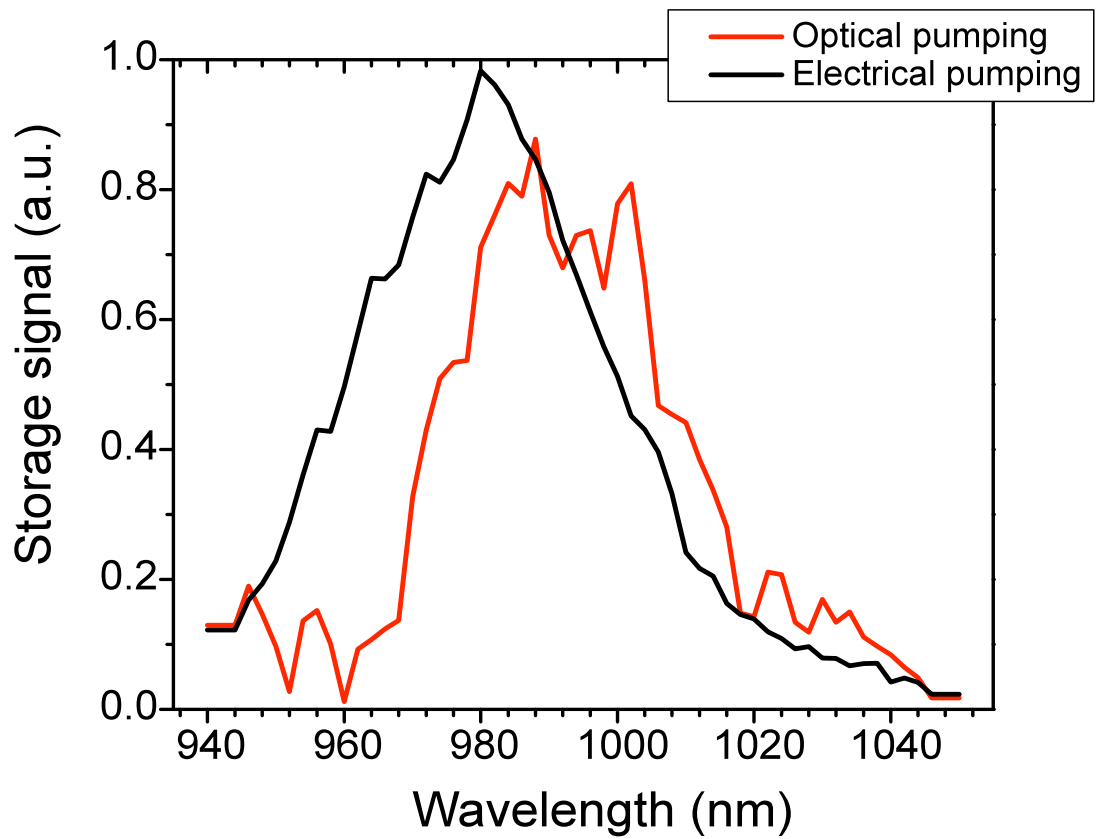

**Figure S5** – Storage spectra recorded with optical (red line) and electrical (black line) pumping for a storage time of 30 μs. For electrical pumping the "write" laser pulse was replaced by an electrical pulse to 3.5V with duration of 1μs. The spectrum is red-shifted for electrical pumping which can be explained by the different values of the electric field during the pumping in the two cases.

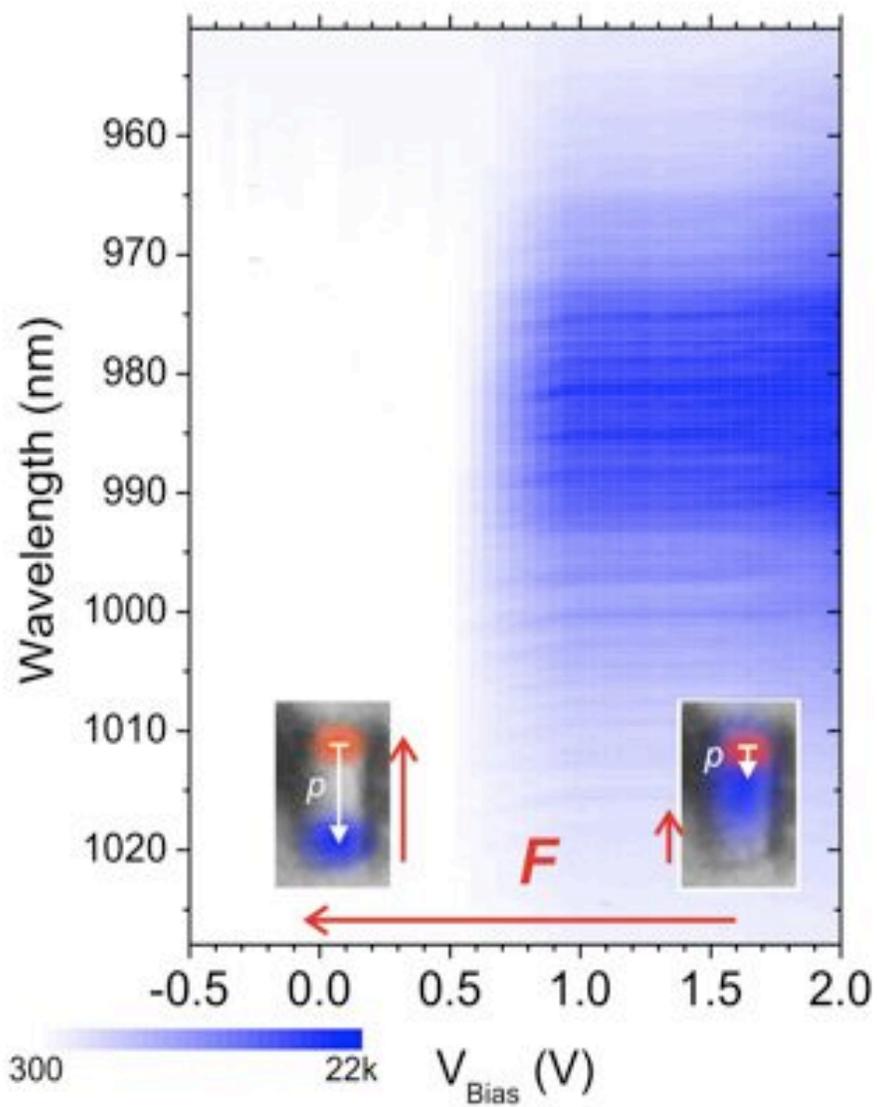

**Figure S6** – Bias-dependent PL spectra of the QP ensemble used for the presented storage experiments. We find the transition to indirect excitons at $V_{Gate} = 0.6$ V.